\documentclass[prd,nofootinbib,12pt]  {revtex4}
\usepackage{amsfonts,amsmath,amscd}
\usepackage{graphicx}
\def\bbone{{\mathchoice {\rm 1\mskip-4mu l} {\rm 1\mskip-4mu l}
{\rm 1\mskip-4.5mu l} {\rm 1\mskip-5mu l}}}

\begin{document}
\small
\title{%
 AdS--Maxwell superalgebra and supergravity}
\author{R. Durka}
\email{rdurka@ift.uni.wroc.pl}\affiliation{Institute for Theoretical
Physics, University of Wroc\l{}aw, Pl.\ Maksa Borna 9, Pl--50-204
Wroc\l{}aw, Poland}
\author{J. Kowalski-Glikman}
\email{jkowalskiglikman@ift.uni.wroc.pl}\affiliation{Institute for
Theoretical Physics, University of Wroc\l{}aw, Pl.\ Maksa Borna 9,
Pl--50-204 Wroc\l{}aw, Poland}
\author{M. Szcz\c{a}chor}
\email{misza@ift.uni.wroc.pl} \affiliation{Institute for Theoretical
Physics, University of Wroc\l{}aw, Pl.\ Maksa Borna 9, Pl--50-204
Wroc\l{}aw, Poland}

\date{\today}

\begin{abstract}
In this paper we derive the Anti de Sitter counterpart of the
super-Maxwell algebra presented recently by Bonanos et.\ al. Then we
gauge this algebra and derive the corresponding supergravity theory,
which turns out to be described by the standard $N=1$ supergravity
lagrangian, up to topological terms.
\end{abstract}

\maketitle

The Maxwell algebra \cite{Bacry:1970ye}, \cite{Schrader:1972zd} (for
recent discussion and references see e.g., \cite{Bonanos:2008ez}) is
an extension of Poincar\'e algebra consisting of the translational
and Lorentz generators $\mathcal{P}_a$ and $\mathcal{M}_{ab}$,
respectively appended with six additional generators
$\mathcal{Z}_{ab}$, forming an antisymmetric Lorentz tensor
satisfying the relation
\begin{equation}\label{1}
    [\mathcal{P}_a, \mathcal{P}_b] = i \mathcal{Z}_{ab}\, ,
\end{equation}
with
\begin{equation}\label{2}
    [\mathcal{M}_{ab},\mathcal{Z}_{cd}]=-i(\eta_{ac}\mathcal{Z}_{bd}+\eta_{bd}\mathcal{Z}_{ac}-\eta_{ad}\mathcal{Z}_{bc}-\eta_{bc}\mathcal{Z}_{ad})
\end{equation}
and $[\mathcal{Z}_{ab},\mathcal{Z}_{cd}]=0$,
$[\mathcal{Z}_{ab},\mathcal{P}_{c}]=0$. In the paper
\cite{Bonanos:2009wy} the authors derive an interesting
supersymmetric $N=1$ extension of this algebra, with two Majorana
supercharges.

In the recent paper \cite{Durka:2011nf} we extended the the Maxwell
algebra (\ref{1}), (\ref{2}) to the AdS-Maxwell one (see also
\cite{Lukierski:2010dy}, \cite{Soroka:2011tc}) and we derived a
dynamical theory resulting from its gauging by using the framework
of constrained BF theories \cite{Starodubtsev:2003xq},
\cite{Smolin:2003qu}, \cite{Freidel:2005ak}. We find that theory
obtained by this procedure is just the Einstein-Cartan theory with
the additional Holst  action term and that the Maxwell field, being
the gauge field associated with the generators $\mathcal{Z}_{ab}$,
appears only in the topological term that does not influence the
dynamics of the theory. This theory differs therefore from the one
discussed in \cite{deAzcarraga:2010sw}; the reason being that in the
latter the Maxwell symmetry was not implemented at the level of the
construction of the action.

In this paper, following the method developed in our earlier work
\cite{Durka:2009pf}, we will construct the N=1 supersymmetric
extension of the model of \cite{Durka:2011nf}. In the first step of
the construction we must find the $N=1$ AdS-Maxwell superalgebra
being the supersymmetric counterpart of the AdS-Maxwell algebra
\cite{Lukierski:2010dy}, \cite{Soroka:2011tc}, \cite{Durka:2011nf}
with the generators $\mathcal{P}_{a}$, $\mathcal{M}_{ab}$, and $
\mathcal{Z}_{ab}$ satisfying the following commutational relations
\begin{align}\label{3}
[\mathcal{P}_{a},\mathcal{P}_{b}]&=i(\mathcal{M}_{ab}-\mathcal{Z}_{ab})\,
,\nonumber\\
[\mathcal{M}_{ab},\mathcal{M}_{cd}]&=-i(\eta_{ac}\mathcal{M}_{bd}+\eta_{bd}\mathcal{M}_{ac}-\eta_{ad}\mathcal{M}_{bc}-\eta_{bc}\mathcal{M}_{ad})
,\nonumber\\
[\mathcal{M}_{ab},\mathcal{Z}_{cd}]&=-i(\eta_{ac}\mathcal{Z}_{bd}+\eta_{bd}\mathcal{Z}_{ac}-\eta_{ad}\mathcal{Z}_{bc}-\eta_{bc}\mathcal{Z}_{ad}),\\
[\mathcal{Z}_{ab},\mathcal{Z}_{cd}]&=-i(\eta_{ac}\mathcal{Z}_{bd}+\eta_{bd}\mathcal{Z}_{ac}-\eta_{ad}\mathcal{Z}_{bc}-\eta_{bc}\mathcal{Z}_{ad}),\nonumber\\
[\mathcal{M}_{ab},\mathcal{P}_c]&=-i(\eta_{ac}\mathcal{P}_{b}-\eta_{bc}\mathcal{P}_{a}),\nonumber\\
[\mathcal{Z}_{ab},\mathcal{P}_{c}]&=0\,.\nonumber
\end{align}
The AdS-Maxwell superalgebra, being a supersymmetric extension of
(\ref{3}) contains two supersymmetric generators $Q_\alpha$ and
$\Sigma_\alpha$, both being Majorana spinors with the following
(anti) commutational rules (this algebra in a slightly different
form was derived previously in  \cite{Bonanos:2010fw})
\begin{align}\label{4}
[\mathcal{M}_{ab}, Q_\alpha] &= - \frac{i}2\, (\gamma_{ab}\,
Q)_\alpha\,,\nonumber\\
 [\mathcal{M}_{ab}, \Sigma_\alpha] &=  -\frac{i}2\, (\gamma_{ab}\, \Sigma)_\alpha\,, \nonumber\\
  [\mathcal{Z}_{ab}, Q_\alpha] &=-\frac{i}2\, (\gamma_{ab}\,
  \Sigma)_\alpha\,,\nonumber\\
   [\mathcal{Z}_{ab}, \Sigma_\alpha] &=-\frac{i}2\, (\gamma_{ab}\, \Sigma)_\alpha\,, \nonumber\\
    [\mathcal{P}_a, Q_\alpha]&= -\frac{i}{2}\, \gamma_a\,( Q_\alpha-
    \Sigma_\alpha)\,,\nonumber\\
     [\mathcal{P}_{a}, \Sigma_\alpha] &=0,\\
\{Q_\alpha, Q_\beta\}& =-\frac{i}{2}(\gamma^{ab})_{\alpha\beta}\,\mathcal{M}_{ab}+
i(\gamma^a)_{\alpha\beta}\, \mathcal{P}_a\, ,\nonumber\\
\{Q_\alpha,\Sigma_\beta\}
&=-\frac{i}{2}(\gamma^{ab})_{\alpha\beta}\,\mathcal{Z}_{ab}\,,\nonumber\\
\{\Sigma_\alpha, \Sigma_\beta\}
&=-\frac{i}{2}(\gamma^{ab})_{\alpha\beta}\,\mathcal{Z}_{ab}\nonumber\,.
\end{align}
By the Wigner-In\"on\"u contraction of the algebra (\ref{3}) with
rescaled generators
 $\mathcal{P}_a \rightarrow a\, \mathcal{P}_a$,
$\mathcal{Z}_{ab} \rightarrow a^2\, \mathcal{Z}_{ab}$ and going with
$a$ to infinity we obtain the standard Maxwell algebra. As for the
supersymmetric extension (\ref{4}), we rescale $Q \rightarrow
a^{1/2}\, Q$ and $\Sigma \rightarrow a^{3/2}\, \Sigma$ to obtain the
Maxwell superalgebra of \cite{Bonanos:2009wy}.

Let us now turn to gauging the AdS-Maxwell superalgebra (\ref{3}),
(\ref{4}). To this end we write down a gauge field, valued in this
superalgebra
\begin{equation}\label{5}
    \mathbb{A}_{\mu}=\frac{1}{2}\omega^{ab}_{\mu}\mathcal{M}_{ab}+\frac{1}{\ell}e^a_{\mu}\mathcal{P}_a+\frac{1}{2}h_{\mu}^{ab}\mathcal{Z}_{ab}+\kappa\bar\psi^\alpha_{\mu} Q_\alpha+\tilde{\kappa}\bar\chi^\alpha_{\mu}\Sigma_\alpha
\end{equation}
In this formula $\ell$ is a scale of dimension of length necessary
for dimensional reason, because the tetrad $e^a_{\mu}$ is
dimensionless. Similarly $\kappa$ and $\tilde{\kappa}$ are scales of
dimension $\mbox{length}^{-1/2}$ included so as to compensate for
the dimension of the spinor fields. As we will see below these
scales are related. The components of the curvature of connection
$\mathbb{A}_{\mu}$
\begin{equation}\label{6}
    \mathbb{F}_{\mu\nu}=\partial_\mu\mathbb{A}_{\nu}-\partial_\nu\mathbb{A}_{\mu}-i[\mathbb{A}_{\mu},\mathbb{A}_{\nu}]\,
\end{equation}
can be written as
\begin{equation}\label{7}
    \mathbb{F}_{\mu\nu}=\frac12\, F^{(s)}_{\mu\nu}{}^{ab}\, \mathcal{M}_{ab}+ F^{(s)}_{\mu\nu}{}^{a}\, \mathcal{M}_{a}+ \frac12\, G_{\mu\nu}^{(s)ab}\,\mathcal{Z}_{ab}+  \bar{\mathcal F}_{\mu\nu}^\alpha Q_\alpha+ \bar{\mathcal G}_{\mu\nu}^\alpha \Sigma_\alpha\,
\end{equation}
where the supercurvatures are given by
\begin{align}
F^{(s)}_{\mu\nu}{}^{ab} &=F_{\mu\nu}^{ab}-\kappa^2\,
\bar\psi_\mu\gamma^{ab}\psi_\nu\, ,\nonumber\\
    F^{(s)}_{\mu\nu}{}^a &=F_{\mu\nu}^{a}+\kappa^2\, \bar\psi_\mu\gamma^{a}\psi_\nu\,
    ,\label{8}
\\
    G^{(s)}_{\mu\nu}{}^{ab} &=G_{\mu\nu}^{ab}-\tilde{\kappa}\kappa\, (\bar\psi_\mu\gamma^{ab}\chi_\nu+\bar\chi_\mu\gamma^{ab}\psi_\nu)-\tilde{\kappa}^2\, \bar\chi_\mu\gamma^{ab}\chi_\nu\,
    ,\nonumber
\end{align}
with the bosonic curvatures
\begin{align}
F_{\mu\nu}^{ab} &=R^{ab}_{\mu\nu}+\frac{1}{\ell^2}( e^a_\mu e^b_\nu- e^a_\nu e^b_\mu)\, ,\nonumber\\
\ell F_{\mu\nu}^{a}     &=D^\omega_\mu e^a_\nu -D^\omega_\nu
e^a_\mu\, ,
\label{9}\\
G_{\mu\nu}^{ab} &=D^\omega_\mu h^{ab}_\nu -D^\omega_\nu
h^{ab}_\mu-\frac{1}{\ell^2}( e^a_\mu e^b_\nu- e^a_\nu e^b_\mu)
+(h^{ac}_\mu h^{\quad b}_{\nu\,c}-h^{ac}_\nu h^{\quad b}_{\mu\,c})\,
    .\nonumber
\end{align}
Notice that the curvature $\ell F_{\mu\nu}^{a}$ is nothing but the
torsion $T_{\mu\nu}^{a}$.

With the help of covariant derivative defined to be
\begin{equation}\label{10}
     \mathcal D_{\mu}\psi_\nu=
\partial_{\mu}\psi_\nu+\frac{1}{4}\omega^{ab}_\mu\,\gamma_{ab}\,\psi_\nu+\frac{1}{2\ell}
e^{a}_{\mu}\,\gamma_{a}\,\psi_\nu=\mathcal
D^\omega_{\mu}\psi_\nu+\frac{1}{2\ell}
e^{a}_{\mu}\,\gamma_{a}\,\psi_\nu\, .
\end{equation}
we can write down the fermionic curvatures in a compact form as
follows
\begin{align}\label{11}
{\mathcal G}_{\mu\nu}&=\tilde\kappa \Big((\mathcal D^\omega_\mu\chi_\nu -\mathcal D^\omega_\nu\chi_\mu) +\frac{1}{4}(h^{ab}_\mu\gamma_{ab}\chi_\nu-h^{ab}_\nu\gamma_{ab}\chi_\mu) \nonumber\\
&+\frac{\kappa}{4\tilde\kappa}(h^{ab}_\mu\gamma_{ab}\psi_\nu-h^{ab}_\nu\gamma_{ab}\psi_\mu)-\frac{1}{2\ell}\frac{\kappa}{\tilde\kappa}\left(e_\mu^a\, \gamma_a\psi_\nu -e_\nu^a\, \gamma_a\psi_\mu\right)\Big)\nonumber\\
{\mathcal F}_{\mu\nu} &=\kappa \left(\mathcal D^\omega_\mu\psi_\nu
-\mathcal D^\omega_\nu\psi_\mu +\frac{1}{2\ell}\left(e_\mu^a\,
\gamma_a\psi_\nu -e_\nu^a\, \gamma_a\psi_\mu\right)\right)\,.
\end{align}

Having these building blocks we can proceed to the construction of
the action of the AdS-Maxwell supergravity. To this end we
generalize the construction presented in \cite{Durka:2009pf}
including an additional 2-form fermionic field $\mathcal{C}^\alpha$
associated with the supercharge $\Sigma_\alpha$. The action reads
\begin{align}
64\pi\mathcal{L} &=\left( B_{\mu\nu}^{IJ} F^{(s)}_{\rho\sigma\,IJ} -
\frac\beta2\, B_{\mu\nu}^{IJ} B_{\rho\sigma\,IJ}
-\frac\alpha4\epsilon_{abcd} B_{\mu\nu}^{ab}B_{\rho\sigma}^{cd}
\right)\epsilon^{\mu\nu\rho\sigma}\nonumber\\
&+\left( C_{\mu\nu}^{ab} G^{(s)}_{\rho\sigma\,ab} - \frac\rho2\,
C_{\mu\nu}^{ab}C_{\rho\sigma\,ab}-\frac\sigma4\epsilon_{abcd}
C_{\mu\nu}^{ab}
C_{\rho\sigma}^{cd}\right)\epsilon^{\mu\nu\rho\sigma}\nonumber\\
&+ \left(\beta\,
C_{\mu\nu}^{ab}B_{\rho\sigma\,ab}+\frac\alpha2\epsilon_{abcd}
C_{\mu\nu}^{ab}
B_{\rho\sigma}^{cd}\right)\epsilon^{\mu\nu\rho\sigma}\nonumber\\
&+4\,\left( \bar{\mathcal B}_{\mu\nu}{\mathcal F}_{\rho\sigma}-
\frac\beta2\,\bar{\mathcal B}_{\mu\nu}{\mathcal B}_{\rho\sigma}-
\frac\alpha2\,\, \bar{\mathcal B}_{\mu\nu}\gamma^5{\mathcal
B}_{\rho\sigma} \right)\epsilon^{\mu\nu\rho\sigma}\nonumber\\
&+4\,\left( \bar{\mathcal C}_{\mu\nu}{\mathcal G}_{\rho\sigma}-
\frac\rho2\,\bar{\mathcal C}_{\mu\nu}{\mathcal C}_{\rho\sigma}-
\frac\sigma2\,\, \bar{\mathcal C}_{\mu\nu}\gamma^5{\mathcal
C}_{\rho\sigma} \right)\epsilon^{\mu\nu\rho\sigma}\,
\nonumber\\
&+4\,\left(\frac{\beta}{2}\,\bar{\mathcal C}_{\mu\nu}{\mathcal
B}_{\rho\sigma}+\frac{\beta}{2}\,\bar{\mathcal B}_{\mu\nu}{\mathcal
C}_{\rho\sigma}+  \frac{\alpha}{2}\, \bar{\mathcal
C}_{\mu\nu}\gamma^5{\mathcal B}_{\rho\sigma} + \frac{\alpha}{2}\,
\bar{\mathcal B}_{\mu\nu}\gamma^5{\mathcal C}_{\rho\sigma}
\right)\epsilon^{\mu\nu\rho\sigma}\,.\label{12}
\end{align}
The bosonic part of this action coincides with the action of
AdS-Maxwell gravity derived in \cite{Durka:2011nf}, while the action
(\ref{12}) with $\mathcal{C}=\mathcal{G}=0$ is just the $N=1$
supergravity action in the constrained BF formalism constructed in
\cite{Durka:2009pf}.

Solving the algebraic field equations for the fermionic two form
fields we find
\begin{align}
\mathcal{B}-\mathcal{C}&=\frac{1}{\alpha^2+\beta^2}\left(\beta\bbone
-\alpha\,\gamma^5\, \right)\mathcal{F}\,, \nonumber\\
 \mathcal{C}&=\frac{(\rho-\beta)\bbone-(\sigma-\alpha)\,\gamma^5}{ (\sigma-\alpha)^{2}+(\rho-\beta)^{2}} \Big( \mathcal{G}+\mathcal{F}
 \Big)\, ,\label{13}
\end{align}
which after substituting back to the fermionic part of the action
(\ref{12}) gives
\begin{align}
16\pi \mathcal
L^{f}&=\epsilon^{\mu\nu\rho\sigma}\frac{\alpha}{(\alpha^2+\beta^2)}\,\bar{\mathcal{F}}_{\mu\nu}\left(
\frac{{\beta}\bbone
-{\alpha}\gamma^5}{2{\alpha}}\right)\,\mathcal{F}_{\rho\sigma} \nonumber\\
&+\epsilon^{\mu\nu\rho\sigma}\frac{(\sigma-\alpha)}{(\sigma-\alpha)^2+(\rho-\beta)^2}\,(\bar{\mathcal{G}}_{\mu\nu}+\bar{\mathcal{F}}_{\mu\nu})\left(
\frac{(\rho-\beta)\bbone
-(\sigma-\alpha)\gamma^5}{2(\sigma-\alpha)}\right)
\,(\mathcal{G}_{\rho\sigma}+\mathcal{F}_{\rho\sigma}) \label{14}
\end{align}
Similarly for the bosonic part of the action we get (see
\cite{Durka:2011nf} for details)
\begin{align}\label{15}
16\pi \mathcal
L^{b}&=\epsilon^{\mu\nu\rho\sigma}\left(\frac{1}{\beta}
F^{(s)a4}{}_{\mu\nu}  F_{a4}^{(s)}{}_{\rho\sigma}+\frac{1}{4}
M^{abcd} F_{ab}^{(s)}{}_{\mu\nu}
F_{cd}^{(s)}{}_{\rho\sigma}\right)\nonumber\\ &+
\epsilon^{\mu\nu\rho\sigma}\,\frac{1}{4}N^{abcd}
\left(G_{ab}^{(s)}{}_{\mu\nu}+F_{ab}^{(s)}{}_{\mu\nu}\right)
\left(G_{cd}^{(s)}{}_{\rho\sigma}+F_{cd}^{(s)}{}_{\rho\sigma}\right)
\end{align}
with
\begin{align}
M^{abcd}&=\frac{\alpha}{(\alpha^2+\beta^2)}( \gamma\,
\delta^{abcd}-\epsilon^{abcd}) \,,\nonumber\\
N^{abcd}&=\frac{(\sigma-\alpha)}{(\sigma-\alpha)^2+(\rho-\beta)^2}\left(\frac{\rho-\beta}{\sigma-\alpha}\delta^{abcd}
-\epsilon^{abcd} \right)\label{15a}
\end{align}
Let us also recall that the parameters of the model $\alpha$,
$\beta$, and $\ell$ are related to the physical coupling constants:
Newton's constant $G$, cosmological constant $\Lambda$,  and Immirzi
parameter $\gamma$ as follows
\begin{equation}\label{15b}
\frac{\Lambda}{3}=-\frac{1}{\ell^2},\quad     \alpha =
\frac{G\Lambda}{3}\frac{1}{(1+\gamma^2)}, \quad \beta =
\frac{G\Lambda}{3}\frac{\gamma}{(1+\gamma^2)} , \quad
\gamma=\frac{\beta}{\alpha}\,.
\end{equation}

Before going further let us pause here for a moment to discuss local
supersymmetry transformations that leave invariant the action being
the sum of the fermionic (\ref{14}) and bosonic (\ref{15}) parts.
The gauge transformation of the connection $\mathbb{A}_{\mu}$
 are defined to be
\begin{equation}\label{16}
    \delta\mathbb{A}_{\mu}=\partial_\mu\Theta-i[\mathbb{A}_{\mu},\Theta]
\end{equation}
with
\begin{equation}\label{17}
    \Theta=\frac{1}{2}\lambda^{ab}\mathcal{M}_{ab}+\xi^a \mathcal{P}_a+\frac{1}{2}\tau^{ab}\mathcal{Z}_{ab}+\bar\epsilon^\alpha Q_\alpha+\bar\zeta^\alpha\Sigma_\alpha\,.
\end{equation}
Substituting (\ref{17}) into (\ref{16}), decomposing the result and
comparing with (\ref{5}) we find for the local supersymmetry
transformations with parameter $\epsilon$
\begin{align}\label{18}
    \delta_\epsilon e_{\mu}^{a} &= -\ell\kappa\,\bar\epsilon\, \gamma^{a}\,\psi\nonumber\\
\delta_\epsilon \omega_{\mu}^{ab} &= \kappa\,\bar\epsilon\, \gamma^{ab}\,\psi\nonumber\\
\delta_\epsilon h_{\mu}^{ab} &= \tilde\kappa\,\bar\epsilon\, \gamma^{ab}\,\chi\\
\delta_\epsilon \bar{\psi}_{\mu} &= \frac{1}{\kappa}(\mathcal D^{\omega}_\mu\bar\epsilon-\frac{1}{2l}e^a_\mu\bar\epsilon\gamma_a)\nonumber\\
\delta_\epsilon \bar{\chi}_{\mu} &= \frac{1}{\tilde\kappa}(
-\frac{1}{4}h^{ab}_\mu\bar\epsilon\gamma_{ab}+\frac{1}{2l}e^a_\mu\bar\epsilon\gamma_a)\nonumber
\end{align}
and with the parameter $\zeta$
\begin{align}\label{19}
\delta_\zeta h_{\mu}^{ab} &= \bar\zeta \gamma^{ab} (\kappa\psi +\tilde\kappa\chi)\nonumber\\
\delta_\zeta \bar{\chi}_{\mu} &= \frac{1}{\tilde\kappa}\mathcal
D^{(\omega+h)}_\mu\bar\zeta\, ,
\end{align}
where the covariant derivative $\mathcal D$ is defined in
(\ref{10}).

Similarly we can find the transformation rules for the curvatures,
to wit
\begin{align}\label{20}
 \delta_\epsilon F_{}^{(s)a4} &= -\bar\epsilon \gamma^{a} {\mathcal F}_{}\nonumber\\
\delta_\epsilon F_{}^{(s)ab} &= \bar\epsilon \gamma^{ab} {\mathcal F}_{}\nonumber\\
\delta_\epsilon G_{}^{(s)ab} &= \bar\epsilon \gamma^{ab} {\mathcal G}_{}\\
\delta_\epsilon \bar{\mathcal F}_{} &= -\frac{1}{4}\bar\epsilon \gamma^{ab}F^{(s)}_{ab}-\frac{1}{2\ell}\bar\epsilon\gamma_a F^{(s)a4}\nonumber\\
\delta_\epsilon \bar{\mathcal G}_{} &= -\frac{1}{4}\bar\epsilon
\gamma^{ab}G_{ab}^{(s)}+\frac{1}{2\ell}\bar\epsilon\gamma_a\nonumber
F^{(s)a4}\, ,
\end{align}
and
\begin{align}\label{21}
\delta_\zeta G_{}^{(s)ab} &= \bar\zeta \gamma^{ab} ({\mathcal F}_{}+{\mathcal G}_{})\nonumber\\
\delta_\zeta \bar{\mathcal G}_{} &= -\frac{1}{4}\bar\zeta
\gamma^{ab}(F^{(s)}_{ab}+G^{(s)}_{ab})\, .
\end{align}
Using (\ref{20}) and (\ref{21}) one can check, using the {\em1.5}
formalism, that the action is indeed invariant under the action of
both these local supersymmetries\footnote{More precisely, the
variation of the action is proportional to super-torsion, which
vanishes in the {\em1.5} formalism.}. The action is of course also
invariant under the bosonic symmetries: the local Lorentz and
Maxwell leave it invariant, and details can be found in
\cite{Durka:2011nf}.

Having convinced ourselves that the action is invariant let us try
to simplify it. Indeed we expect a lot of cancelations taking place.
As we know from \cite{Durka:2011nf} the Maxwell gauge field
$h_\mu{}^{ab}$ appears in the bosonic action in the topological
terms and, as a consequence of this, its superpartner $\chi$ should
disappear from the action as well. Let us check if this is indeed
what is happening.

To see this let us first notice that the curvatures $F^{(s)}$ and
$\mathcal F$ have exactly the same form as in the $N=1$ AdS
supergravity discussed in \cite{Durka:2009pf}, so that we must only
consider the  $F^{(s)}+G^{(s)}$ and $\mathcal F+\mathcal G$ terms in
the lagrangians (\ref{14}), (\ref{15}). These terms have the form
\begin{align}\label{22}
G_{\mu\nu}^{(s)ab}+F_{\mu\nu}^{(s)ab}&=R^{ab}_{\mu\nu}(\omega+h)-\Big(\kappa\bar\psi_\mu+\tilde
\kappa\bar\chi_\mu\Big)\gamma^{ab}\Big(\kappa\psi_\nu+\tilde
\kappa\chi_\nu\Big)\nonumber\\
    \mathcal{G}_{\mu\nu}+\mathcal{F}_{\mu\nu}&=\mathcal D^{(\omega+h)}_\mu \Big(\kappa\psi_\nu+\tilde \kappa\chi_\nu\Big)-\mathcal D^{(\omega+h)}_\nu \Big(\kappa\psi_\mu+\tilde
    \kappa\chi_\mu\Big)\,.
\end{align}
 Using this and
\begin{align}
&    \bar{\mathcal{F}}_{\mu\nu}\left( \frac{\bbone{\beta} -\gamma^5{\alpha}}{2{\alpha}}\right)\,\mathcal{F}_{\rho\sigma}\,\epsilon^{\mu\nu\rho\sigma}=\nonumber\\
&=4\frac{\kappa^2}{2}
\mathcal (D^{}_\mu \bar\psi_\nu)\,(\gamma\bbone -\gamma^5)\mathcal (D^{}_\rho \psi_\sigma)\,\epsilon^{\mu\nu\rho\sigma}\nonumber\\
    &=\frac{\kappa^2}{4}\bar\psi_\mu\,(\gamma\bbone -\gamma^5)\, \left(\gamma_{ab} \,F^{ab}_{\nu\rho}+\gamma_a\frac{2}{\ell}T_{\nu\rho}^a\right)\,\psi_\sigma\,\epsilon^{\mu\nu\rho\sigma}\nonumber\\
    &+\kappa^2\,\bar\psi_\mu\,\left(\frac{1}{\ell^2}\gamma^5\gamma_{ab}\,e^a_\nu\, e^b_\rho+\frac{2}{\ell}\gamma^5\gamma_{a}\,e^a_\nu\,\mathcal D^\omega_\rho\right)\psi_\sigma\,\epsilon^{\mu\nu\rho\sigma}\label{22a}\\
    &+\mbox{\em total derivative}\nonumber
\end{align}
 after some straightforward but tedious calculations one
 can bring the Lagrangian to the following form
\begin{align}\label{23}
  16\pi \mathcal L&=-\left(\frac{\kappa^2}{G}\,\bar\psi_\mu\,\gamma^5\gamma_{ab}\,e^a_\nu\, e^b_\rho+ \frac{2\kappa^2\ell}{G}\,\bar\psi_\mu\,\gamma^5\gamma_{a}\,e^a_\nu\,\mathcal D^\omega_\rho\psi_\sigma\right)\,\epsilon^{\mu\nu\rho\sigma}\nonumber\\
&-\bar\psi_\mu\,\left(\frac{1}{4\beta} \frac{2\kappa^2}{\ell}\gamma_{a}\,T_{\nu\rho}^a+\frac{2\kappa^2\ell}{4G}\,(\gamma\bbone -\gamma^5)\, \gamma_a\,T_{\nu\rho}^a\right)\,\psi_\sigma\,\epsilon^{\mu\nu\rho\sigma}\nonumber\\
&-\frac{1}{4\beta} \left(\frac{1}{\ell^2}T_{\mu\nu}^{a}\,T_{\rho\sigma\,a}+\kappa^4 \,\bar\psi_\mu\gamma^{a}\psi_\nu \, \bar\psi_\rho\gamma_{a}\psi_\sigma\right)\,\epsilon^{\mu\nu\rho\sigma}\\
&+\frac{1}{16} M_{abcd} \left(F_{\mu\nu}^{ab}\,F_{\rho\sigma}^{cd}+\kappa^4\, \bar\psi_\mu\gamma^{ab}\psi_\nu\, \bar\psi_\rho\gamma^{cd}\psi_\sigma\right)\,\epsilon^{\mu\nu\rho\sigma}\nonumber\\
&+\frac{1}{16}N_{abcd}\left(\kappa\bar\psi_\mu+\bar
\kappa\bar\chi_\mu\right)\gamma^{ab}\left(\kappa\psi_\nu+\bar
\kappa\chi_\nu\right) \left(\kappa\bar\psi_\rho+\bar
\kappa\bar\chi_\rho\right)\gamma^{cd}\left(\kappa\psi_\sigma+\bar
\kappa\chi_\sigma\right)\,\epsilon^{\mu\nu\rho\sigma}\nonumber\\
&+\mbox{\em total derivative}\nonumber
\end{align}
 Making use of  Fierz identities one can check that the last line in (\ref{23})
vanishes identically along with other four-fermion terms and there
are some simplifications in the second line. Notice that in this way
there is no trace  of $\chi$ in the bulk Lagrangian anymore. Indeed,
after some cancelations all the $\chi$-dependent terms can be
combined into a total derivative.

The   Lagrangian reduces therefore to the final form
\begin{align}\label{24}
&   16\pi \mathcal L=\left(
\frac{1}{16} M_{abcd} \,F_{\mu\nu}^{ab}\,F_{\rho\sigma}^{cd}-\frac{1}{4\beta\ell^2} T_{\mu\nu}^{a}\,T_{\rho\sigma\,a}\right)\,\epsilon^{\mu\nu\rho\sigma}\nonumber\\
&-\left(\frac{\kappa^2}{G}\,\bar\psi_\mu\,\gamma^5\gamma_{ab}\,e^a_\nu\, e^b_\rho+\frac{2\kappa^2\ell}{G}\,\bar\psi_\mu\,\gamma^5\gamma_{a}\,e^a_\nu\,\mathcal D^\omega_\rho\psi_\sigma\right)\,\epsilon^{\mu\nu\rho\sigma}\nonumber\\
&+\frac{\kappa^2\ell}{2\gamma G}\,\bar\psi_\mu
\gamma_{a}\,\psi_\nu\,T_{\rho\sigma}^a \,\epsilon^{\mu\nu\rho\sigma}
+\mbox{\em total derivative}\,,
\end{align}
which, up to the total derivative terms being the superymmetric
generalization of those found in the bosonic case
\cite{Durka:2011nf}, is just the $N=1$ supergravity Lagrangian, cf.\
\cite{Durka:2009pf} if we set $\kappa^2=\frac{4\pi G}{\ell}$.
Although these topological terms do not change the bulk field
equations they may influence the asymptotic charges in an
interesting way. We will investigate this in a separate paper.

This concludes our construction, in which we showed that the gauged
theory of the AdS-Maxwell supersymmetry is somehow trivial, reducing
just to the standard $N=1$ supergravity. In view of
\cite{Durka:2011nf} this result is hardly surprising, since in that
paper we found that in the bosonic case the gauge field of Maxwell
symmetry $h_\mu{}^{ab}$ appears similarly  only through a
topological term.

\section*{ACKNOWLEDGEMENTS} The work of J.\ Kowalski-Glikman is supported in part by the
grant 182/N-QGG/2008/0, the work of R.\ Durka is supported by the
National Science Centre grant  N202 112740, and the work of R.\
Durka and M.\ Szczachor was supported by the European Human Capital
Programme.

\end{document}